# Ultimate limit of field confinement by surface plasmon polaritons

*Jacob B Khurgin*

*Johns Hopkins University Baltimore MD 21218 USA*

jakek@jhu.edu

We show that electric field confinement in surface plasmon polaritons propagating at the metal/dielectric interfaces enhances the loss due to Landau damping and which effectively limits the degree of confinement itself. We prove that Landau damping and associated with it surface collision damping follow directly from Lindhard formula for the dielectric constant of free electron gas Furthermore, we demonstrate that even if all the conventional loss mechanisms, caused by phonons, electron-electron, and interface roughness scattering, were eliminated, the maximum attainable degree of confinement and the loss accompanying it would not change significantly compared to the best existing plasmonic materials, such as silver.

**Introduction**

Steady advances in nanofabrication made in the last decade had inspired research in nano-plasmonics, a field that carries many exciting promises in various areas of technology such as sub-wavelength imaging, sensing, nano-scale optical interconnects and active devices [1,2]. In one way or another, these promises are all hinged upon the ability to concentrate optical field into the sub-wavelength dimensions that is a salient feature of surface plasmon polaritons (SPP's), whose nature is a combination of electro-magnetic field with charge waves of free carriers in metal (or semiconductor). When the dimensions are reduced way below wavelength the magnetic field is greatly diminished (static limit), and the energy which is normally stored in the form of magnetic energy is instead stored in the form of kinetic energy of carriers (kinetic inductance) which makes sub-wavelength oscillation mode sustainable. Unfortunately and inevitably, the free carrier oscillations dissipate energy at a very high rate, of the order of $\gamma_m \sim 10^{14} s^{-1}$ in noble metals and $10^{13} s^{-1}$ in highly doped semiconductors. As a result, the losses in the SPP which are significantly sub-wavelength (in case of propagating SPP's it means the SPP propagation constant β much larger than wave-vector in dielectric $k_d$) are always on the scale of $\gamma_m$ independent on the shape and exact size as long as it is significantly sub wavelength [3]. Due to these high losses the promises of nanoplasmonics, which include miniature efficient sources and detectors of radiation, nano-scale optical interconnects, super-resolution imaging, and others has not been fully realized yet. [4]

The reason the optical losses in metals are high is unfortunately the same one that makes support charge oscillations at optical frequencies sustainable in the first place. High free carrier density (Fermi level high in the conduction band) requisite to maintain high plasma frequency also assures that the density of states into which these electrons can scatter is also very high, with strong scattering following from this unfortunate fact. In optical range the metal loss is caused by more than one mechanism [5] – scattering by phonons, carrier-carrier Umklapp processes, residual intra-band absorption and scattering on surface imperfections can all contribute on more or less equal scales, hence their reduction is not simple. Nevertheless, some strides in that direction are being made. Most obviously, reduction of metal surface roughness achieved with epitaxial growth have yielded reduced loss in Ag [6], while using Al in place of Ag [7] reduced parasitic interband absorption in the blue part of spectrum. The phonon-assisted absorption can be reduced, but not eliminated by going to cryogenic temperatures because the spontaneous phonon emission is possible even at absolute zero, while the temperature-independent Umklapp scattering may be somewhat mitigated in the materials with different (less spherical) shapes of Brillouin zone, although, once again, the numerous efforts with materials as diverse as ITO and TiN [8] so far have not shown substantial improvement over ubiquitous noble metals.

Nevertheless, the hope is alive, that sooner or later, a novel material with negative ε and substantially lower losses in the optical range will emerge, and in anticipation of these developments, it is worthwhile to estimate their practical impact, i.e. what would be the maximum attainable degree of field concentration if all scattering-related loss in the metal in the optical range were essentially eliminated. Obviously, this question has been raised before. For instance, it has been long known that since SPP is a combination of field and electronics oscillation, maximum SPP wave vector $K_{max}$ cannot possibly exceed Fermi wave vector $k_F \sim 1.2 \times 10^8$ cm$^{-1}$ for noble metals, i.e. restricting the degree of field concentration to about 5 A. This is also the scale at which electron tunneling in the nanogap in the plasmonic dimers commences [9] that further assures that electric field cannot be confined to sub-nanometer dimensions. Yet this limit, usually referred to as "quantum limit" has not been achieved experimentally, and more recently a different explanation which has put the limit of file concentration in the range of a few nanometers has been put forward. The explanation was based on the non-locality phenomenon [10], or, in simpler terms on special dispersion of dielectric constant ε(***k***). Using hydrodynamic model of nonlocality Mortensen et al [11] have shown that when characteristic dimensions of the system become comparable with characteristic length $l_c = v_F/\omega$, where $v_F$ is a Fermi velocity and $\omega$ is a frequency, the nonlocal effects prevent the light from tight confinement and broaden resonances, particularly for the dimer structures. Larkin and Stockman [12] pointed that spatial dispersion effects limit the resolution of the plasmonic superlens to about 5nm in the visible range, i.e. comparable to a few $l_c$. They also pointed out that spatial dispersion is intimately related to Landau damping –i.e. direct absorption of electromagnetic waves by electrons below the Fermi level. Landau damping in the nanoscale metallic objects can also be interpreted as quantum confinement effect or as absorption assisted by surface collisions as explained by Kreibig and Volmer [13], who introduced the phenomenological expression for this process as a "surface collision scattering rate" $\gamma_s \sim v_F/a$ where $v_F$ is Fermi velocity $a$ is the characteristic dimension of the metallic object. According to this phenomenological treatment the additional damping is caused by the limited physical dimension of the system and is the result of restriction of mean free path of electrons.

However, to the best of our knowledge [14-15], there is no unified treatment of nonlocality (i.e. spatial dispersion of real part of ε) and Landau damping (i.e. spatial dependence of the imaginary part of ε) which would allow one to provide an unambiguous answer on which of two phenomena exerts stronger influence on SPP properties. In this work we develop this unified treatment and show that **it is Landau damping, i.e. loss induced by the field confinement itself that is responsible for most dramatic limitations in plasmonics. We show that in case of propagating SPP on single metal-dielectric interface, the limitations arise due to the field (and not electron) confinement and thus are not influenced by the metal dimensions.** Landau damping is shown to become important when characteristic dimensions are on the scale of 10nm which is at least an order of magnitude larger than $l_c$, and, finally, and to some degree surprisingly, it is shown that total elimination of all other loss mechanisms in metals will not yield noticeable improvement in field confinement and power dissipation compared to the best plasmonic materials of today.

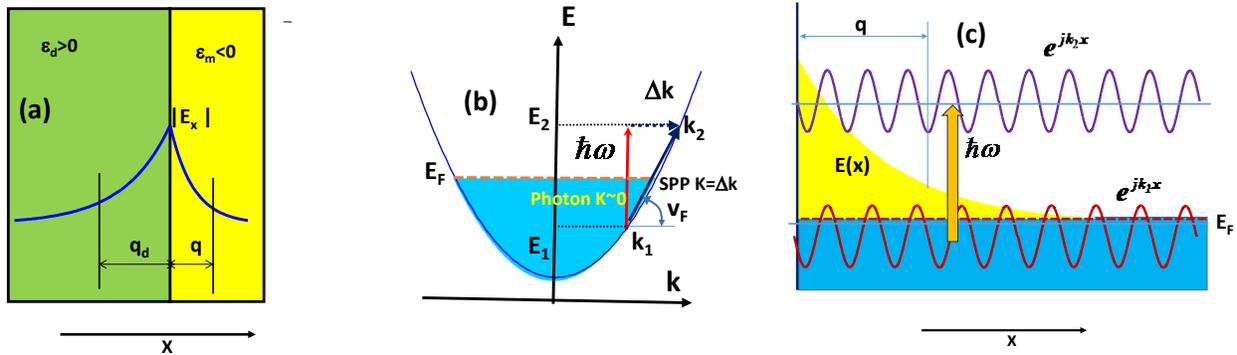

**Fig.1 (a)** Sketch of the fields in the propagating SPP **(b)** Phonon or impurity assisted absorption of a photon with a wavevector K~0 and a direct absorption of SPP with large wavevector K (Landau Damping) **(c)** wavefunctions involved in the absorption of SPP

**Damping rate of SPP due to surface collisions**

We start with the case of SPP propagating at the interface of metal $\varepsilon_m < 0$ and dielectric $\varepsilon_d > 0$ as shown in Fig1.a with propagating constant $\beta$ and two components of electric field, normal

$$E_x = \begin{cases} E_0 e^{-qx} \cos(\beta z - \omega t) & x > 0 \\ \dfrac{\varepsilon_m}{\varepsilon_d} E_0 e^{q_d x} \cos(\beta z - \omega t) & x < 0 \end{cases} \quad (1)$$

that is subjected to damping by surface collisions, and also the tangential one

$$E_z = \begin{cases} \dfrac{q}{\beta} E_0 e^{-qx} \sin(\beta z - \omega t) & x > 0 \\ \dfrac{q_m}{\beta} E_0 e^{q_d x} \sin(\beta z - \omega t) & x < 0 \end{cases} \quad (2)$$

which is not damped by surface collisions. For the absorption in the metal to take place, the electron with energy $E_1$ and wavevector $k_1$ below the Fermi level make a transition to a state above it with energy $E_2 = E_1 + \hbar\omega$ and wavevector $k_2$ resulting in a wavevector mismatch $\Delta k$ that makes transition forbidden (Fig.1b) The value of mismatch is on the scale of $\Delta k_c \sim \omega/v_F \sim l_c^{-1}$ where Fermi velocity $v_F$ is about $1.4 \times 10^8 cm/s$ for noble metals, i.e. for the 1eV photon energy the wave-vector mismatch is about $1nm^{-1}$ and is thus much larger than wavevector of electromagnetic wave. As a result, absorption usually involves additional act of scattering, due to phonons, electron-electron scattering, or imperfections that occur at the rate $\gamma_0$ which occurs on the scale of a few tens of femtoseconds in most plasmonic metals. At first glance, in the SPP both $\beta$ and $q$ are much less than $1nm^{-1}$ one should not expect `much absorption. But if one takes a look at the normal direction, one can see that electric field contains all kind of wavevectors – indeed by taking the one dimensional Fourier transform of (1) we obtain the power spectrum of it

$$|E(K)|^2 = E_0^2 \frac{q/\pi}{K^2 + q^2} \quad (3)$$

Thus the fraction of power of the wave with wavevectors exceeding $\Delta k_c$ is on the order of $F(K > \Delta k_c) \approx (2/\pi)(q/\Delta k_c)$. The penetration depth $L_p \sim (2q)^{-1}$ of the SPP is typically on the scale of a few tens of nanometers, hence a few percent of the SPP energy does wavevector sufficient for Landau damping to take place. Landau damping is shown schematically in Fig1.b as direct "diagonal" transition between the states $k_1$ and $k_2$ caused by the electromagnetic wave with wave-vector $K = \Delta k_c$. As a result of Landau damping imaginary part of the dielectric constant of the metal, describing the energy loss by electromagnetic wave (and hence by SPP) to the individual electronic excitations will increase. The phenomenological "surface collision rate" $\gamma_s$ introduced by Kreibig modifies the expression for the effective dielectric constant of the metal as

$$\varepsilon_{eff}''(\omega) = \frac{\omega_p^2 (\gamma_0 + \gamma_s)}{\omega^3} \quad (4)$$

where $\gamma_0$ is the momentum relaxation rate due to phonons, electron-electron scattering and defects, usually referred to as bulk scattering rat., The surface collision rate may be introduced phenomenologically [10,13], or quantum-mechanically [16], but it is simply added to the expression for dielectric constant using Matthiessen rule, and is not directly connected to nonlocality.

We now evaluate the energy damping rate of SPP polaritons that is due to localization of the field in the vicinity of the metal interface with dielectric. As one can see from the Fig1.c the two free electron wavefunctions $\psi_{1(2)} = L^{-1/2} e^{i\mathbf{k}_\parallel \cdot \mathbf{r}_\parallel} e^{ik_{x1(2)}}$ ($L$ being the normalization length) are orthogonal and optical transition between them is not allowed. But since the electric field is localized according to (1), the square of the interaction Hamiltonian between two states can be found as

$$|H_{12}|^2 = \frac{e^2 |\langle 1|\mathbf{p}\cdot\mathbf{A}|2\rangle|^2}{m^2} = \frac{e^2 \hbar^2 k_{x1} k_{x2} E_0^2}{4m^2 \omega^2 L^2 (\Delta k^2 + q^2)} \tag{5}$$

Note that the fact that electrons get reflected by the surface is not reflected in (5) – the existence of transition is strictly due to the confinement of the field. Since both states are close to Fermi energy, i.e. $\hbar\omega \ll E_F$, we can make two important approximations $\hbar\omega \approx \hbar v_{Fx} \Delta k$ where $v_{Fx}$ is the transverse component of velocity on the Fermi surface, and, furthermore $\hbar^2 k_{1x} k_{2x} / 2m \approx m v_{Fx}^2 / 2m$, which upon substitution into (5) yields

$$|H_{12}|^2 = \frac{e^2 v_{Fx}^4 E_0^2}{4\omega^4 L^2 \left(1 + q^2 v_{Fx}^2 / \omega^2\right)} \tag{6}$$

We now invoke Fermi Golden rule to evaluate the total rate of the field induced upward transitions from the state 1, $R_1 = 2\pi |H_{12}|^2 L\rho_x(E_F)$, where $\rho_x(E_F) = \left(2\pi\hbar |v_{Fx}|\right)^{-1}$ is one-dimensional density of the final states, evaluated under consideration that neither spin nor direction of propagation change as the transition takes place. We thus obtain

$$R_1 = \frac{e^2 |v_{Fx}^3| E_0^2}{4\hbar^2 \omega^4 L \left(1 + q^2 v_{Fx}^2 / \omega^2\right)} \tag{7}$$

Next we perform summation over all the states 1 inside Fermi sphere from which transitions into unoccupied states can take place. That involves integration over the Fermi surface and then multiplying by $\hbar\omega$ as well as the normalization length $L$ to obtain the surface rate of excitation of hot electrons per unit of area. Integration over the Fermi surface is simply multiplication by the three-dimensional density of sates $\rho_{3D} = m^2 v_F / \pi^2 \hbar^3$ and averaging over the angles which yields $|v_{Fx}^3| = v_x^3 / 4$, so we obtain

$$\frac{dN_{2D}}{dt} = \frac{e^2 m^2 v_F^4 E_0^2}{16\pi^2 \hbar^4 \omega^3 (1 + q^2/\Delta k_c^2)} \tag{8}$$

Since $q \ll \Delta k_c$ we can write the expression for the energy loss at surface

$$-\frac{dU_{2D}}{dt} = \frac{e^2 m^2 v_F^4 E_0^2}{16\pi^2 \hbar^3 \omega^2} \tag{9}$$

where $U_{2D}$ is the time-averaged two-dimensional density of kinetic energy of electrons in the SPP which can be found as the integral of three-dimensional density of energy

$$U_{2D}(x) = \int_0^\infty \frac{1}{4}\left(\frac{\partial(\omega\varepsilon_m)}{\partial\omega} - \varepsilon_b\right) E_x^2(x) dx = \frac{N_e}{8q} \frac{e^2 E_0^2}{m\omega^2} \tag{10}$$

where $\varepsilon_b$ is the part of dielectric constant due to bound electrons, and the electron density in a parabolic band is $N_e = k_F^3/3\pi^2 = m^3 v_F^3/3\pi^2\hbar^3$, hence

$$U_{2D} = \frac{1}{24\pi^2 q} \frac{e^2 m^2 v_F^3 E_0^2}{\hbar^3 \omega^2} \tag{11}$$

Comparison of (9) and (11) immediately results in the energy relaxation rate

$$\frac{1}{U_{2D}} \frac{dU_{2D}}{dt} = -\frac{3}{2} q v_F = -2\gamma_s \tag{12}$$

where $\gamma_s$ is the momentum scattering rate due to surface collisions

$$\gamma_s = \frac{3}{4} q v_F \tag{13}$$

that enters into expression (4) for the dielectric constant. This expression is not much different from the one in previous works, where the scattering rate for a nanoparticle with radius R is $\gamma_s = A v_F / R$ but in this work we have obtained this expression using full quantum mechanical derivation. The fact that our coefficient *A* is less than values for nanoparticles is easy to explain by the fact that our problem is one dimensional, hence not every electron contributes to the surface absorption.

**It is very important to stress that the "surface collision damping rate" obtained by us does not require the electrons to be confined, or even reflected– simple confinement of light on the scale of penetration length** $L_{pen} = 1/2q$ **is sufficient to overcome momentum conservation rules and cause substantial absorption even if the electrons are considered to**

**be free**. In other words the confinement of light is what causes the damping, hence the proper name for $\gamma_s$ should be "Landau damping rate", or, better, "time-of-flight" broadening. Nevertheless, we shall use "surface collisions damping" term to conform to the existing literature.

**Surface collision damping as a non-local effect**

Next we demonstrate that result (13) can also be obtained by using a simple, phenomenological picture of simply "shifting" the resonance frequency in Drude formula to account for the possibility of "diagonal in k-space" transition as shown Fig 1b. In the classical Drude formula, obtained in the long wavelength limit of Lindhard approximation the energy difference between two states involved in the optical transition is zero, but for the electromagnetic wave with the finite wavevector K the resonance is shifted by roughly $\Delta\omega = \alpha v_F K$ where $\alpha$ is on a scale of unity. Using Klimontovich-Silin-Lindhard approximate formula [17] for the dielectric constant of the metal one can obtain $\alpha = \sqrt{3/5}$

$$\varepsilon(\omega, K) = \varepsilon_b + \frac{3}{K^2 v_F^2}\left[1 - \frac{\omega}{2Kv_F}\ln\frac{\omega + Kv_F}{\omega - Kv_F}\right] \approx \varepsilon_b - \frac{\omega_p^2}{\omega^2 - \frac{3}{5}K^2 v_F^2 + i\omega\gamma_0} \tag{14}$$

where $\gamma_0$ is the "intrinsic" scattering due to phonons, defects and electron-electron scattering, and $\varepsilon_b$ is the interband contribution to dielectric constant ($\varepsilon_b \approx 4.1$ for Ag)  Equation (14) is easy to interpret as modification of Drude formula. For negligibly small wave vectors the resonant energy of the transition between two free electron states is zero, but as wave-vector increases the resonance shifts upward, by roughly $\Delta\omega \approx \sqrt{3/5}Kv_F \ll \omega$ where Landau Damping takes place (the factor of 3/5 can be traced to the averaging over the Fermi surface). As shown in Fig2a the whole dispersion curve shifts towards higher frequencies resulting in small change of the dielectric constant. Separating dielectric constant into the real and imaginary parts one can write

$$\varepsilon_r(\omega, K) = \varepsilon_b - \frac{\omega_p^2}{\omega^2}\frac{K_0^2\left(K_0^2 - K^2\right)}{\left(K_0^2 - K^2\right)^2 + K_0^4\gamma_0^2/\omega^2}$$

$$\varepsilon_i(\omega, K) = \frac{\omega_p^2}{\omega^2}\frac{K_0^4\gamma_0/\omega}{\left[K_0^2 - K^2\right]^2 + K_0^4\gamma_0^2/\omega^2} \tag{15}$$

where $K_0 = \sqrt{5/3}\Delta k_c$ is roughly the wave-vector at which "diagonal" absorption of light (Landau Damping) commences.

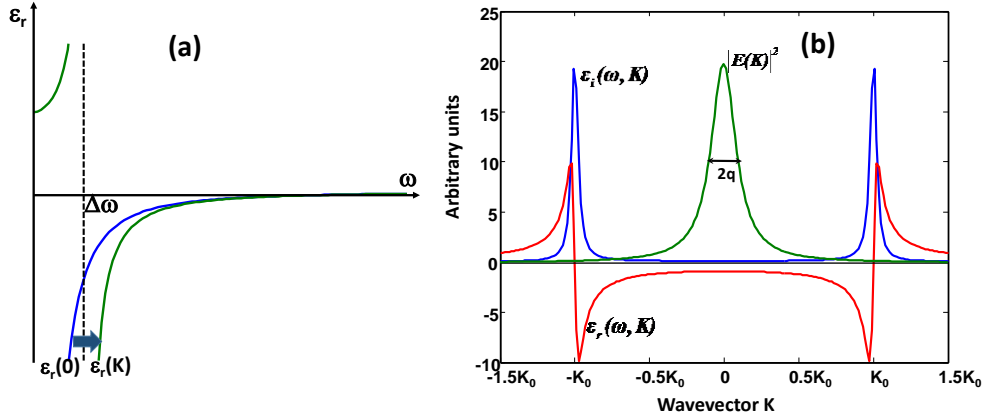

**Fig.2** (a) Sketch of the resonance shift due to non-locality (Eq (14). (b) Effective dielectric constant of SPP can be obtained by overlapping spatial power spectrum of SPP $|E(k)|^2$ with wave-vector dependent dielectric constant (real and imaginary parts)

The plot of wavevector-dielectric constant is shown in Fig.2b for $\omega/\gamma_0 = 20$ and for the imaginary part of dielectric constant it consists of essentially flat response for $K \ll K_0$ followed by the sharp delta-function like Lorentzian peaks near $K = \pm K_0$

$$\varepsilon_i(\omega, K) \approx \begin{cases} \dfrac{\omega_p^2 \gamma_0}{\omega^3} & |K| \ll K_0 \\ \dfrac{\omega_p^2}{\omega^2} \dfrac{K_0^2 \gamma_0 / 4\omega}{[K \mp K_0]^2 + K_0^2 \gamma_0^2 / 4\omega^2} \approx \dfrac{\omega_p^2}{\omega^2} \dfrac{\pi K_0}{2} \delta(K \mp K_0) & K \approx \pm K_0 \end{cases} \quad (16)$$

The effective dielectric constant can be obtained by integrating (16) over the normalized power density spectrum of the electric field inside the metal $|E(K)|^2 = q/\pi(K^2 + q^2)$, also of Lorentzan shape as plotted in Fig 2b for $q = K_0/10$, which immediately gives us

$$\varepsilon_{i,eff}(\omega, q) = \frac{\omega_p^2 \gamma_0}{\omega^3} + \frac{\omega_p^2}{\omega^2} \frac{qK_0}{K_0^2 + q^2} \approx \frac{\omega_p^2 \gamma_0}{\omega^3} + \frac{\omega_p^2 \sqrt{3/5} q v_F}{\omega^3} = \frac{\omega_p^2 (\gamma_0 + \gamma_s)}{\omega^3} \quad (17)$$

where the surface collision broadening is $\gamma_s = \sqrt{3/5} q v_F \approx 0.77 q v_F$ i.e. result that is very close to quantum mechanical derivation (13). Besides providing simple physical interpretation, equation (21) also confirms the Matthiessen's rule of the strength of absorption induced by surface reflection $\gamma_s$ being added to the strength of absorption induced by all other means $\gamma_0$.

If we now turn our attention to the real part of the dielectric constant, then one can see that over the region where the $K \ll K_0$, the power density spectrum $|E(K)|^2$ looks like a delta function, while the integrating over the real part of Lorentzian leads to cancellation, and, as a result $\varepsilon_{r,eff}(\omega) \approx \varepsilon_r(\omega, 0)$. In other words, the nonlocality effects are dominated by the change in imaginary part of the dielectric constant, i.e. surface collision damping (a.k.a. time-of-flight broadening)

**Ultimate limit of field confinement imposed by surface collision damping**

We now what to see how the increased damping due to field confinement manifests itself for the propagating SPP's of Fig1.a. The propagation constant of SPP can be found as

$$\beta(\omega) = k_D(\omega)\sqrt{\frac{\varepsilon_m(\omega)}{\varepsilon_d + \varepsilon_m(\omega)}} \tag{18}$$

where $k_D(\omega) = 2\pi\varepsilon_d^{1/2}\omega/c$ is the wavevector of the free propagating electromagnetic wave in the dielectric with frequency-independent dielectric constant $\varepsilon_d$ and the metal dielectric constant, according to (14) and (17) given by

$$\varepsilon_{\text{eff}}(\omega,q) \approx \varepsilon_b - \frac{\omega_p^2}{\omega^2 + i\omega[\gamma_0 + \gamma_s(q)]} \tag{19}$$

and the collision-induced damping rate being

$$\gamma_s = \frac{3}{4}q(\omega)v_F f_x(\omega) \tag{20}$$

where

$$q(\omega) = \sqrt{\beta^2(\omega) - \varepsilon_m k_D^2} \tag{21}$$

and

$$f_x = \frac{|E_x|^2}{|E_z|^2 + |E_x|^2} = \frac{\beta^2}{\beta^2 + q^2} \tag{22}$$

is a fraction of the energy contained in the normal component of electric field. If we now introduce the frequency of SP resonance, $\omega_{SP} = \omega_p/\sqrt{(\varepsilon_b + \varepsilon_d)}$ and normalize the frequencies to it, $\tilde{\omega} = \omega/\omega_{SP}$, and introduce effective index of SPP as $\tilde{\beta} = \beta/k_D$ and normalized decay constant $\tilde{q} = q/k_D$ we obtain from (18) and (19)

$$\tilde{\beta}(\tilde{\omega}) = \sqrt{\frac{\varepsilon_b(\tilde{\omega}^2 - 1) - \varepsilon_d + i\varepsilon_b\tilde{\omega}Q_0^{-1} + i\varepsilon_b\tilde{\omega}^2 Q_s^{-1}}{(\varepsilon_b + \varepsilon_d)(\tilde{\omega}^2 - 1 + i\tilde{\omega}Q_0^{-1} + i\tilde{\omega}^2 Q_s^{-1})}} \tag{23}$$

where $Q_0 = \omega_{SP}/\gamma_0$ and $Q_s(\tilde{\omega}) = 4c/3\varepsilon_d^{1/2} v_F f_x \tilde{q}_r$ and $\tilde{q}_r(\tilde{\omega})$ is the real part of normalized decay constant. Since $Q_s(\tilde{\omega})$ itself is a function of $\tilde{\beta}(\tilde{\omega})$ equation (23) can be solved self-consistently, by iterations, but prior to that a few important observations can be made.

First, the impact of surface damping would become noticeable when $\gamma_s \sim \gamma_0$, which will happen not far from the SP resonance, hence when $Q_s \sim Q_0$, and since in the vicinity of SPP resonance $\tilde{q} \approx \tilde{\beta}$ and $f_x \approx 1/2$ we obtain the value of effective index at which the surface dumping mast be taken into account,

$$\tilde{\beta}_s \approx \frac{8c}{3\varepsilon_d^{1/2} v_F Q_0} \tag{24}$$

If we consider the combination of Ag and GaN ( $\varepsilon_d^{1/2} = 2.3$, $\gamma_0 = 3.2 \times 10^{13} s^{-1}$, $\omega_{SP} = 4.5 \times 10^{15} s^{-1}$; $Q_0 \approx 140$; [18] ) we obtain $\tilde{\beta}_s \approx 1.6$ while for GaN –Au combination ($\gamma_0 = 1.2 \times 10^{14} s^{-1}$; $Q_0 \approx 43$; [18]) we obtain $\tilde{\beta}_s \approx 5$. As expected, it is for good metal, like silver that surface collision role becomes important early on, while for the less perfect metal, like gold the influence of surface collision does not become important until much later.

Second, one can use (23) to find the ultimate value of the effective index, and hence confinement, of SPP which could have been obtained in the hypothetical "ideal" metal with $\gamma_0 = 0$, i.e. free of defects, phonon scattering, electron-electron interaction, and residual interband absorption. Obviously, such metal does not exist, however it is useful to see what kind of improvement can be achieved by reducing the loss in the metal. By inserting $\tilde{\omega} = 1$ and $Q_0^{-1} = 0$; into (23) we obtain

$$\tilde{\beta}_{max} = \sqrt{\frac{\varepsilon_d i Q_s}{(\varepsilon_b + \varepsilon_d)}} \approx \sqrt{i \frac{\varepsilon_d^{1/2}}{(\varepsilon_b + \varepsilon_d)} \frac{8c}{3v_F \tilde{\beta}_{max,r}}} \tag{25}$$

Therefore we obtain a rather simple expression for the maximum effective index (real part) attainable with the "ideal" metal

$$\tilde{\beta}_{max,r} \approx \left( \frac{\varepsilon_d^{1/2}}{\varepsilon_b + \varepsilon_d} \frac{4c}{3v_F} \right)^{1/3} \tag{26}$$

Then for a wide variety of dielectrics transparent in the visible and near IR with refractive indices between 1.5 and 3, $4 < \tilde{\beta}_{max,r} < 4.5$, so one arrives at a rather surprising result – one can reduce the wavelength of the SPP propagating on the metal-dielectric interface by no more than a factor of 4.5 relative to the plane wave propagating in dielectric, no matter how low is the loss in the bulk metal.

## Numerical Results

We now calculate the dispersion curves (23) for the SPP propagating on the boundary between the metal with silver-like dispersion ($\varepsilon_b = 4.1$, $\hbar\omega_{SP} = 9.3eV$;) and GaN $\varepsilon_d^{1/2} = 2.3$ resulting in SPP resonance near 415nm. When it comes to the scattering constant in the metal we shall consider 5 different cases:

A. The best bulk silver with bulk damping constant $\gamma_0 = 3.2 \times 10^{13} s^{-1}$ and no surface collision damping taken into account $\gamma_s = 0$

B. The best bulk silver with bulk damping constant $\gamma_0 = 3.2 \times 10^{13} s^{-1}$ with surface collision damping taken into account

C. "Dirty silver" with bulk damping constant $\gamma_0 = 1.2 \times 10^{14} s^{-1}$ similar to that of gold and no surface collision damping taken into account $\gamma_s = 0$. The reason for using "dirty silver" instead of gold is that one cannot observe interface SP resonance in gold in combination with any dielectric due to high interband absorption, but to see how the surface collision damping affects metals with reasonably high bulk loss is important.

D. "Dirty silver" with bulk damping constant $\gamma_0 = 1.2 \times 10^{14} s^{-1}$ similar to that of gold with surface collision damping taken into account.

E. "Ideal metal" with no bulk damping $\gamma_0 = 0$, with surface collision damping taken into account.

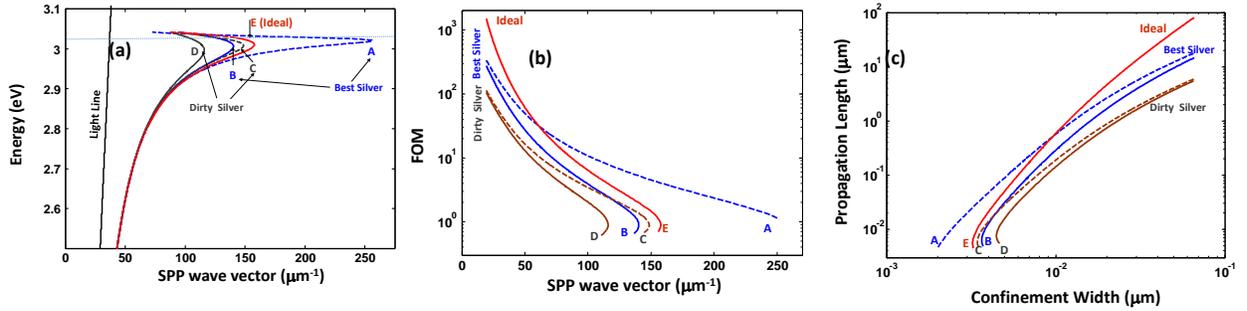

**Fig.3** (a) dispersion curves (b) Figure of Merit vs wavevector (c) Propagation length vs. Confinement width of SPP on the interface between Ga-N and Ag-like metal for the following cases: (A) The best bulk silver with bulk damping constant $\gamma_0 = 3.2 \times 10^{13} s^{-1}$ and no surface collision damping taken into account $\gamma_s = 0$ (B) Same with surface collision damping taken into account (C) "Dirty silver" with bulk damping constant $\gamma_0 = 1.2 \times 10^{14} s^{-1}$ similar to that of gold and no surface collision damping taken into account $\gamma_s = 0$. (D) Same with bulk damping constant $\gamma_0 = 1.2 \times 10^{14} s^{-1}$ similar to that of gold with surface collision damping taken into account. (E) "Ideal metal" with no bulk damping $\gamma_0 = 0$, with surface collision damping taken into account.

The results are shown in Fig. 3a, as well as the light line representing propagation of plane electromagnetic wave in GaN. As expected, not taking into account surface damping for the best silver (curve A) leads to very large propagation constant, with effective index exceeding 7, but once surface collisions damping have been included (curve B) the propagation constant is reduced almost two-fold. If we now consider the more realistic silver, full of defects due to surface deposition process, whose damping rate is comparable to gold, the curves without (C) and with (D) surface collision damping, the latter's impact is less significant, although still prominent. But it is the curve (E) which is most telling – if one starts with the best available silver and then hypothetically gets rid of all damping processes, being it defects, phonons or electron-electron interaction, then, even if the surface is atomically smooth the increase of the attainable propagation constant $\beta_{\max,r}$ (and hence the degree of confinement $q$) will be only about 12%, and the maximum effective index will not exceed roughly 4.4, just as predicted by (26). That means the minimum confinement depth in the normal direction $d_{\min,x} = (q_d^{-1} + q^{-1})/2$ will not be less then roughly $\lambda_D / 25 n_d$ where, $n_d = \varepsilon_d^{1/2}$. Furthermore, if we now consider the imaging using SPP's, the superlens [19,20] and apply the analysis of [21], the minimum spot size (confinement in lateral direction) achievable in this configuration would be $d_{\min,x} \approx 2\pi / \beta_{\max,r} \approx \lambda / 4.4 n$, i.e, close to the result obtained in [12].

To show how surface collision damping affects losses we also plot in Fig.3b the figure of merit, defined as the ratio $FOM = \text{Re}(\beta)/2\text{Im}(\beta)$, or as one can say, the phase shift accumulated over one absorption length. Once again we can see that for the high quality silver (curves A and B) the impact of surface collisions becomes important at large wavevectors with almost an order of magnitude difference, achieved at $\beta_r = 150 \mu m^{-1}$ while for the higher loss metal (curves C and D) the difference is somewhat less. But the most important is the fact that for the "ideal" metal FOM improves by only a factor of two relative to the "best silver"

Yet another way to show the effect of surface collision damping is to plot the propagation length $L_{prop} = 1/2\text{Im}(\beta)$ vs. the field penetration width in dielectric $L_{con} = 1/2\text{Re}(q_d)$ as demonstrated in Fig.3c. Reducing the bulk losses $\gamma_0$ helps to increase propagation length by about an order of magnitude for the confinement of wider than 50nm, but for tighter confinement, as surface collision damping becomes dominant, and curves B,D, and E get close to each other, getting rid of all bulk losses results in only marginal increase of the propagation length

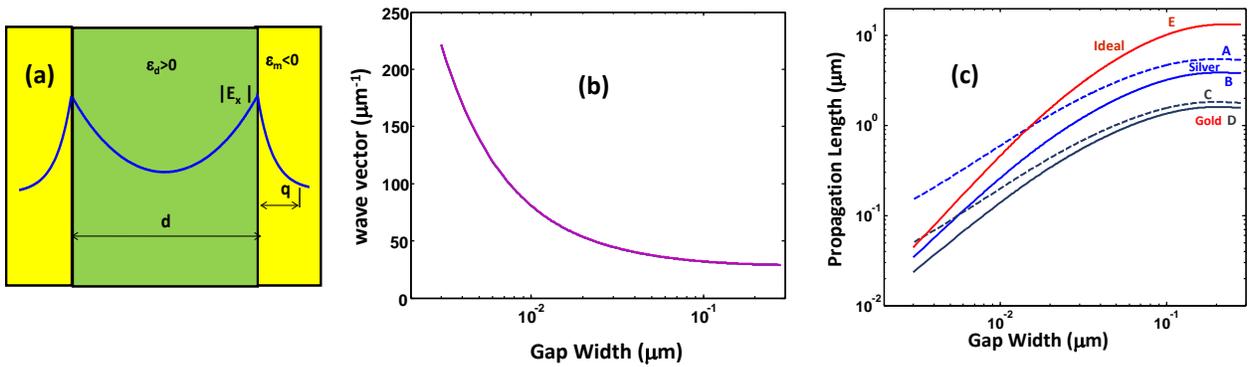

**Fig.3** (a) sketch of gap SPP (b) Propagation length vs. Confinement width of gap SPP with Ga-N core and metal cladding (A) The best bulk silver with bulk damping constant $\gamma_0 = 3.2 \times 10^{13} s^{-1}$ and no surface collision damping taken into account $\gamma_s = 0$ (B) Same with surface collision damping taken into account (C) Gold with bulk damping constant $\gamma_0 = 1.2 \times 10^{14} s^{-1}$ similar to that of gold and no surface collision damping taken into account $\gamma_s = 0$. (D) Same with bulk damping constant $\gamma_0 = 1.2 \times 10^{14} s^{-1}$ similar to that of gold with surface collision damping taken into account. (E) "Ideal metal" with no bulk damping $\gamma_0 = 0$, with surface collision damping taken into account.

To further emphasize this point we consider the case of gap plasmon [22], Fig.4a with GaAs core and metal cladding, made either of Ag or Au. Once again we consider the same 5 cases as above, except cases C and D now refer to gold, as well as to "dirty silver". The dispersion curves for all 5 cases, shown in Fig.4b are essentially, identical, but the amount of loss differ dramatically, as shown in Fig. 4c where the propagation length is plotted versus the gap width. As one can see inclusion of surface collision damping greatly reduces propagation length for silver (curves A and B) and somewhat less than that for gold. The most significant observation to be made from this figure is that hypothetically avoiding all bulk loss in the "ideal" metal (curve E) does increase propagation length for the weakly confined gap plasmons with gap size over 100nm, but for the tightly confined ones, with gap size less than 50nm the effect is marginal. It appears that surface collision damping alone makes propagation length shorter than 1μm, which makes gap plasmon impractical for application as, say, interconnect.

**Conclusions**

In this work we have considered the impact of non-local effects on the properties of propagating SPP's. We have shown that the increase in loss is due to the final extent of the optical field, rather than due to collisions with the surface. In other words, surface collision damping/broadening is better described as time-of-flight broadening. We have obtained full quantum-mechanical expression for the damping rate and corresponding change in the imaginary part of dielectric constant. We have confirmed that the increased damping is a non-local effect that follows naturally from the Lindhard theory of the wavevector-dependent dielectric constant and does not have to be introduced phenomenologically. We have shown that nonlocality – engendered change in the imaginary part of dielectric constant exerts much stronger influence on properties of plasmonic structures than the dispersion of the real part.

We then applied the theory to the case of propagating SPP's and have shown that not only surface collision damping increases loss, but it actually prevents the field from being concentrated into the tight regions. As a result, even if the bulk scattering of the metal had been completely eliminated, one would have not be able to concentrate the field into the regions that are substantially tighter than roughly $\lambda/4.5n_d$ i.e exceeding the diffraction limit only by a factor of few. One can show that similarly, in case of dimers, not only the line width will broaden as two nanoparticles will get closer [14], but the mode itself will expand. This can be simply understood using coupled mode analysis [23] in which the gap mode in dimer consist of superposition of multipole modes. Since higher order modes are strongly confined near surface, they are damped relative to lower order modes, and as a result they do not get excited as easily as lower order modes.

These results raise a very important question about the impact of the efforts to reduce bulk loss in metals. It appears that eliminating the defects and imperfections of fabrication process, and, hypothetically, reducing phonon and electron-electron scattering will not reduce loss significantly relative to today's best silver for the substantially (factor of few) sub-wavelength structures. There are of course other compelling reasons for looking at different materials, such as cost and compatibility with CMOS processes for integration of Plasmonics with electronics. Furthermore, when it comes to the structures that are not tightly confined such as long range SPP's [Bern] the already relatively low loss can be further reducing using material with a smaller bulk loss. But once the confinement gets really tight surface collision damping makes losses high and nearly independent of bulk losses. It seems that as long as there exist two electronic states separated by photon energy, one occupied and one empty, there will always be a transition between two of them, and hence absorption. The only way to avoid all losses, including Landau damping is to develop materials with narrow conduction bands, such that there is no empty state within $\hbar\omega$ from the bottom of the band [24], so that transition cannot take place by any means. Perhaps that is where the effort to develop low loss materials should be directed.

The author acknowledges fruitful discussions with A. V. Uskov